\documentclass[letter]{aa}

\usepackage{graphicx}
\usepackage{txfonts}
\usepackage{amsmath}
\usepackage{natbib}
\usepackage{xcolor}
\usepackage{hyperref}
\hypersetup{colorlinks=true, linkcolor=black, urlcolor=black, citecolor=blue}
\bibpunct{(}{)}{;}{a}{}{,}

\newcommand{\blfootnote}[1]{%
  \begingroup
  \renewcommand{\thefootnote}{}\footnote{#1}%
  \addtocounter{footnote}{-1}%
  \endgroup
}

\newcommand{\zoa}{ZoA}

\begin{document}

\title{The extragalactic science of the KMOS VVVX-GalCen spectroscopic survey}
\subtitle{III. Discovery of a galaxy cluster at z = 0.089 behind the Galactic plane}
\titlerunning{Spectroscopic confirmation of a galaxy cluster in the Zone of Avoidance}
\authorrunning{Schmidt et al.}

\author{E.~O.~Schmidt, C.~Valotto, C. Villalon, L. D. Baravalle,  J. L. Nilo-Castell\'on,  M. V. Alonso, M.~Soto, M.~G\'omez et al.
\\ \normalsize\textit{(Full author list and affiliations can be found after the references)}}

\institute{}
\date{Received XXX; accepted YYY}

\abstract
{The Zone of Avoidance (\zoa) continues to represent the most significant incompleteness 
in the extragalactic census of the Local Universe. 
It is evident that photometric surveys alone are insufficient to establish whether the galaxy overdensities detected therein are physical systems or projections along the line of sight.}
{
We present the first spectroscopic results of the extragalactic science area of the
KMOS VVVX-GalCen Spectroscopic Survey, demonstrating how near-infrared (NIR) spectroscopy can
reveal and dynamically characterise galaxy clusters hidden by the Milky Way.}
{A photometric overdensity of 19 galaxy candidates was identified in the VVV NIRGC~III
catalogue, centred at $(l,b)=(274.05\degr,+2.85\degr)$ with a mean interstellar extinction
of $A_{K_{\rm s}}=0.2$~mag; 16 of these candidates were observed with VLT/KMOS in the
$HK$ grating and yielded spectra with reliable cross-correlation radial velocities.}
{
A total of 14 galaxies form a compact structure with a mean redshift of $\langle z \rangle =0.089\pm0.001$, which corresponds to a mean radial velocity $\langle cz \rangle = 26\,660 \pm 170$ km~s$^{-1}$, while two are foreground galaxies.
The rest-frame line-of-sight velocity dispersion is $\sigma_v=560^{+110}_{-120}$~km~s$^{-1}$, implying an indicative dynamical mass of $M_{200}=1.9^{+0.7}_{-1.1}\times10^{14}\,M_\odot$ (based on members within the cluster core $\sim0.4\,r_{200}$) and placing the system in the poor-cluster regime. Of this 14 cluster members, 13 also lie within the $2\sigma$ scatter of the fitted near-IR red sequence. The other two galaxies are  foreground objects that lie outside this range.}
{The newly obtained radial-velocity measurements provide the first spectroscopic confirmation of a galaxy cluster identified by the extragalactic science area of the KMOS VVVX-GalCen Spectroscopic Survey. 
The results demonstrate the capability of combining VVVX NIR photometry with KMOS spectroscopy to identify and dynamically characterise galaxy clusters in regions of the \zoa.}


\maketitle

\section{Introduction}
\label{sec:intro}

A reliable map of the mass distribution in the Local Universe requires a
whole-sky galaxy census, yet roughly one-tenth of the sky remains
poorly explored. Along the Galactic plane, interstellar dust, and (at the lowest
latitudes) severe stellar crowding obscure background galaxies, even in the
NIR. This region,  called the Zone of Avoidance (\zoa), intersects with some
of the richest structures in the nearby Universe, including the Great Attractor
\citep{lynden1988} as well as the Norma \citep{Kraan1996} and Vela \citep{kraan2017}
superclusters. Recovering the galaxies hidden behind the Galactic plane is
therefore essential for reconstructing the local mass distribution and velocity field. 

Deep NIR imaging from the VISTA Variables in the V\'ia L\'actea survey
\citep[VVV;][]{minniti2010} and its extension VVVX \citep{Saito2024} has
substantially increased the number of galaxies detected in this region, yielding catalogues
of more than $200000$ galaxy candidates \citep{baravalle2021, Daza2023,
Duplancic2024, Alonso2025}. Using the VVV NIRGC I catalogue, \citet{Soto2022}
identified photometric overdensities that may extend known superclusters into
the obscured region. These NIR surveys have also supported cluster studies behind the
Galactic plane along two complementary lines. VVV/VVVX photometry has
been used to characterise clusters previously known from X-ray observations,
such as a cluster detected with Suzaku \citep{Coldwell2014} and the
Ophiuchus cluster \citep{Galdeano2022}. In addition,
\citet{Baravalle2019} and \citet{Galdeano2023} identified new cluster
candidates from VVV near-infrared (NIR) photometry, which were subsequently
confirmed as physical structures through long-slit spectroscopic observations
with Gemini South. As these cases show, photometry alone cannot
determine whether an apparent overdensity corresponds to a physical structure
or to a chance alignment of galaxies at different distances along a crowded
line of sight. Redshift measurements are needed to turn the 2D
catalogue of galaxies into a 3D view of the galaxy distribution.

The KMOS VVVX-GalCen Spectroscopic Survey \citep[ESO programme ID 116.29EF;][]{Arnaboldi2026} 
is designed to deliver NIR spectroscopy of Galactic and extragalactic sources at low Galactic latitudes, exploiting the multiplexing capability
of the 24 integral field units (IFUs) of 
KMOS at the VLT \citep{Sharples2013}. The survey design, target selection,
observing strategy, data reduction, and redshift-measurement methodology for the
extragalactic science area of the survey are fully
described in the
companion Early Data Release (EDR) paper (Baravalle et al., in prep.),
which releases the first 161 galaxy spectra and validates the end-to-end
pipeline.

In this Letter, we present the first galaxy cluster spectroscopically confirmed by the survey. 
The system, whose brightest cluster galaxy (BCG) is VVVX-J094137.62$-$490302.5, was
identified as a photometric overdensity at $(l,b)=(274.05\degr,+2.85\degr)$ using the VVV NIRGC~III catalogue
\citep{Alonso2025}. It has been confirmed in this work based on
spectroscopy with KMOS.
Section~\ref{sec:data} summarises the sample and observations, 
Sect.~\ref{sec:results} presents the membership, dynamics ,and red sequence (RS), while
Sect.~\ref{sec:disc} discusses the implications. Throughout this work, we employ a flat
$\Lambda$CDM cosmology with $H_0=70$~km~s$^{-1}$~Mpc$^{-1}$, $\Omega_m=0.3$,
and $\Omega_\Lambda=0.7$ \citep{planck2020}.

\section{Sample and observations}
\label{sec:data}

Following the photometric overdensity methodology of \citet{Soto2022}, we
identified a galaxy overdensity in the VVV NIRGC~III catalogue
\citep{Alonso2025} centred at $(l, b) = (274.05\degr,\, 2.85\degr)$, within a
circular region of radius $0.1\degr$ ($\sim6$~arcmin).
We characterised the interstellar extinction across the field from the colour excess of red-clump stars, using the catalogue of \citet{Alonso2026} and following the approach developed by \citet{Gonzalez2012} for the Galactic bulge, converting the resulting colour excess to $K_s$-band extinction via $E(B-V) = A_{K_{\rm s}} / 0.351$. This yields a mean interstellar extinction in this field of $A_{K_s} = 0.2$~mag. 
The 19 photometric candidates span a magnitude range of
$13.19 \leq K_s \leq 15.90$~mag. Visual inspection of the $JHK_s$ colour composite
images confirms that all candidates display smooth, centrally concentrated morphologies
typical of early-type galaxies in the NIR, supporting their identification
as candidate cluster members.

Spectroscopic observations were obtained with KMOS using the $HK$ grating ($14600\le\lambda\le24100~\AA$),  with a spectral sampling of $\approx 4.6$~\AA~pixel$^{-1}$ and a spectral resolution of $R \approx 2000$ (FWHM$_{\rm inst} \approx 150$~km~s$^{-1}$) as part of the survey. The instrument configuration, KMOS Arm Allocation and Observing Block preparation are detailed in the EDR. Observations were carried out on March 12 and April 10, 2026. Of the 19
photometric candidates, 16 were targeted with KMOS and yielded spectra of
sufficient quality to enable reliable radial-velocity measurements, while
3 were not observed due to constraints in the KMOS arm allocation and therefore lack
spectroscopic information. Data reduction followed the end-to-end procedure validated in the EDR: KMOS data were processed with the ESO Data Processing System (EDPS; \citealt{Freudling2024}) using the dedicated KMOS pipeline (v4.5.3), including master calibrations, wavelength calibration, illumination correction, and sky subtraction, with telluric correction performed via \textsc{Molecfit} \citep{Smette2015, Kausch2015} and 1D extraction from each reconstructed cube with \textsc{CARTA} \citep{comrie2026}.

Radial velocities were measured with the classic Fourier-space cross-correlation method \citep{Tonry1979} against 23 late-type (G7-M5) stellar templates from the IRTF Spectral Library \citep{Cushing2005, Rayner2009}. 
We required a Tonry--Davis reliability parameter $R\ge7$ and a consensus among templates. We combined the retained templates as an $R^2$-weighted mean \citep{Tonry1979}.
The full methodology is described in the EDR paper. 
Each cross-correlation function was inspected visually and individually. All 16 observed galaxies yielded a reliable radial velocity. These radial velocities were independently confirmed with KROSS (Obasi et al., 2026 in press.), with radial velocities in very good agreement to within $\sim$ 1\%, providing an additional check on the spectroscopic measurements.
The photometric and spectroscopic properties of the candidate sample are summarised in Table~\ref{tab:cat} in Appendix \ref{sec:appendix}.
Formal errors from this cross-correlation analysis are known to underestimate the true uncertainty \citep[e.g.][]{Collins1995, Zamora2023}, as they capture only the precision of the template matching and are insensitive to systematics from wavelength calibration, sky subtraction, 1D extraction, and other steps of the reduction chain, which are difficult to quantify individually \citep[e.g.][]{Schmidt2018}. In this context, the velocity errors listed in Table~\ref{tab:cat} reflect only this internal cross-correlation precision, that is, the scatter of the radial velocities recovered across templates, and are consistently treated as a lower limit on the global uncertainty (Baravalle et al., in prep.).

\begin{figure}
\centering
\includegraphics[width=\columnwidth]{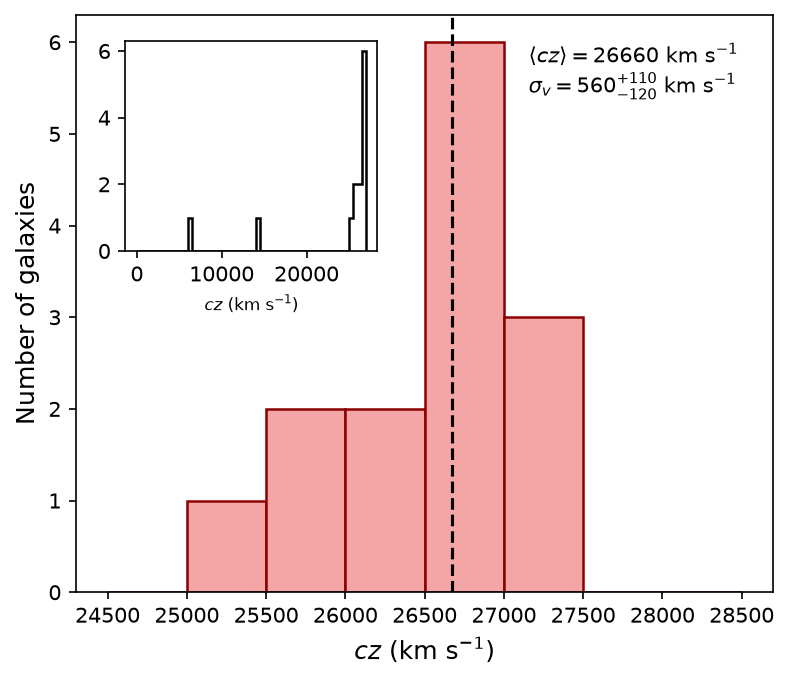}
\caption{Distribution of radial velocities of the 14 cluster
members. The dashed line marks the mean radial velocity of the cluster,
$\langle cz \rangle = 26\,660$ km~s$^{-1}$. The inset shows the full velocity range of the 16 galaxies observed with KMOS.
}
\label{fig:vhist}
\end{figure}

\section{Results}
\label{sec:results}

With the radial velocities obtained in Sect.~\ref{sec:data}, we can assess which galaxies belong to a genuine overdensity, estimate its distance, and derive its dynamical properties. A total of 14 galaxies are concentrated within a narrow velocity interval, $25\,380\le cz \le27\,480$~km~s$^{-1}$, 
defining a prominent peak in the velocity distribution, which is presented in Fig.~\ref{fig:vhist}.
The two remaining galaxies are separated from the main concentration by more than 20 times the velocity dispersion of the system and therefore  unambiguously identified as foreground galaxies. The 14 members confirmed spectroscopically have a mean redshift of  $\langle z \rangle =0.089\pm0.001$, which corresponds to a luminosity distance of $410$~Mpc. The radial velocity has a mean value of $\langle cz \rangle = 26\,660 \pm 170$ km~s$^{-1}$, with a median value of $cz = 26\,810$~km~s$^{-1}$.

The line-of-sight velocity dispersion, computed using the gapper estimator \citep{Beers1990}, which is well suited to samples of this size, is $\sigma_v=560^{+110}_{-120}$~km~s$^{-1}$ in the cluster rest frame ($600$~km~s$^{-1}$ in the observed frame), with confidence intervals derived from $20\,000$ bootstrap resamplings. The full velocity range spanned by the members, $2\,100$~km~s$^{-1}$, is nevertheless consistent with the measured dispersion: a Monte Carlo draw of 14 velocities from a Gaussian with $\sigma_v=560$~km~s$^{-1}$ produces a range this large or larger in $42\%$ of realisations.

Adopting the velocity-dispersion--mass scaling calibrated on cosmological simulations by \citet{Evrard2008}, this velocity dispersion implies an indicative dynamical mass estimate,
\begin{equation}
M_{200}=1.9^{+0.7}_{-1.1}\times10^{14}\,M_\odot ,
\label{eq:mass}
\end{equation}
with $r_{200}\simeq1.2$~Mpc. These values are consistent with a relatively low-mass cluster, placing the system in the poor-cluster regime. However, they should not be taken as a robust measurement of the total cluster mass, since the present observations establish the cluster core, but this is not yet its full spatial extent or richness.

\begin{figure}
\centering
\includegraphics[width=\columnwidth]{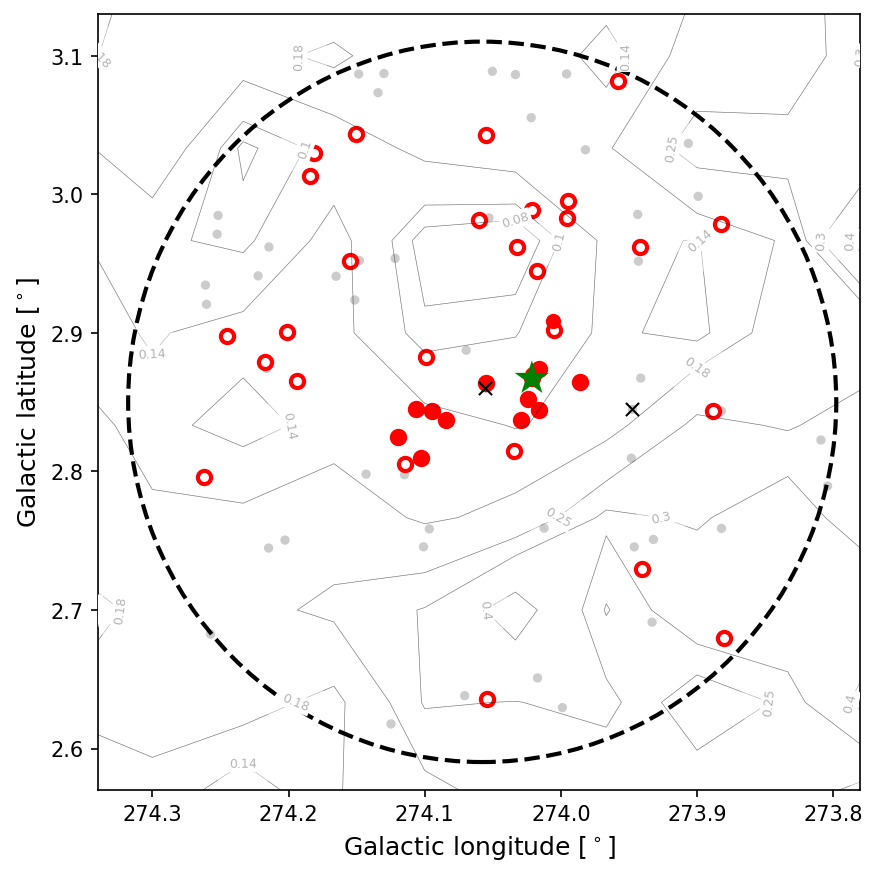}
\caption{Projected distribution of galaxies in Galactic
coordinates. Grey circles are VVVX galaxies in the field; open red circles are
photometric RS candidates without spectroscopy; filled red circles
are the 14 spectroscopically confirmed cluster members; black crosses mark
the two non-member galaxies; and the
green star marks the BCG of the system. Thin contours show the
interstellar extinction, $A_{K_s}$, with values in magnitudes labelled along each
curve. The dashed black circle indicates the projected search radius.}
\label{fig:map}
\end{figure}

Figure~\ref{fig:map} shows the projected distribution of the spectroscopically 
confirmed cluster members. All 14 members lie within $4\farcm6$ of the member centroid, corresponding 
to $\sim0.47$~Mpc or $\sim0.4\,r_{200}$. 
Thus, the spectroscopic sampling is largely confined to the central region of the cluster.
The derived $\sigma_v$ should therefore be interpreted as representative of this region rather than of the cluster as a whole. Consequently, applying a global velocity-dispersion--mass relation may therefore introduce an additional systematic uncertainty in $M_{200}$ that cannot be quantified with the present data.

\begin{figure}
\centering
\includegraphics[width=\columnwidth]{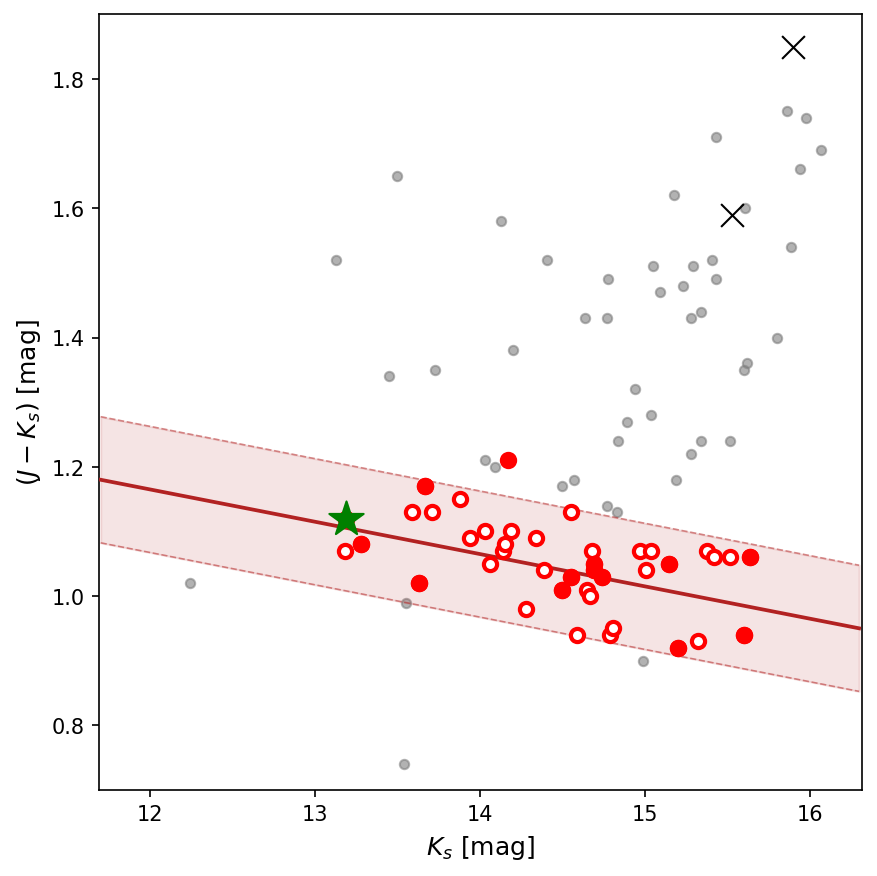}
\caption{NIR CMD. The
RS (solid line, Eq.~\ref{eq:rs}, with its $\pm2\sigma_{\rm RS}$
envelope shaded) is fitted using only the 14 spectroscopically confirmed cluster members. 
Filled red circles indicate galaxies with radial velocities consistent with cluster membership, while open red circles mark galaxies without spectroscopic confirmation that lie within $\pm2\sigma_{\rm RS}$ envelope.
Grey points show all galaxies in the field. The green star marks the BCG.
}
\label{fig:cmd}
\end{figure}

Figure~\ref{fig:cmd} shows the near-IR colour-magnitude diagram (CMD) for all galaxies identified within $(r_{200})$ of the cluster centre. The 19 photometric candidates that motivated the spectroscopic observations are included in this sample, of which 16 were observed with KMOS. The RS \citep{Gladders2000} was fitted using only the 14 spectroscopically confirmed cluster members, excluding the two foreground galaxies via
\begin{equation}
(J-K_{\rm s})=(-0.050\pm0.019)\,K_{\rm s}+(1.764\pm0.269),
\label{eq:rs}
\end{equation}
with an observed scatter $\sigma_{\rm RS}=0.051$~mag. The slope differs from zero at only $2.7\sigma$ and the fit has $r^2=0.40$ over the $2.45$~mag baseline spanned by the 14 galaxies; given the small sample, we attach no physical significance to the detailed form of the fit. Since the RS is defined by the confirmed members themselves, it does not constitute independent evidence for the existence of the cluster.

The fitted RS is subsequently used as a complementary photometric diagnostic for the galaxies without spectroscopic measurements. In particular, objects lying within the ($\pm2\sigma_{\rm RS}$) envelope are consistent with the colour distribution of the spectroscopically confirmed members, but this criterion alone does not establish cluster membership. Conversely, the confirmed member that falls outside this envelope is retained as a member on the basis of its radial velocity.

A search of the literature revealed only one previously published radial velocity in the field, associated with the galaxy KK\,1547, with a value of $17\,476\pm162$~km~s$^{-1}$ \citep{Kraan1995}.
This galaxy corresponds to VVVX-J094115.64$-$485952.9 in the VVV NIRGC~III
catalogue.
However, this galaxy is not among those observed in our KMOS sample and its radial velocity 
places it well in the foreground of the cluster. KK\,1547 is located at the bright end of the CMD ($K_s=12.24$), but its colour lies more than $2\sigma$ below the best-fitting RS. Its position in the CMD therefore provides additional evidence supporting its identification as a foreground galaxy rather than a cluster member.

The three candidates without spectroscopy all fall within the $\pm2\sigma_{\rm
RS}$ envelope of this spectroscopically defined RS and, thus, they can be considered as plausible additional members (although this is only a photometric expectation, not a spectroscopic one. Unlike the candidates described just before, these three were never part of the fit, so their falling on the sequence is a genuine prediction rather
than a fitted outcome. This is consistent with the confirmed
core of the cluster sitting inside a photometric overdensity that might extend
beyond the 16 galaxies observed so far. This is the argument we make in the context of the
projected distribution given in Fig.~\ref{fig:map}. The CMD is
dominated by the RS, with a scarcity of blue galaxies, as expected for an evolved
cluster; the NIR selection of the parent catalogue nevertheless
favours red objects and, thus, this cannot be understood as a measurement of the blue fraction.

\section{Discussion and conclusions}
\label{sec:disc}

We present the first galaxy cluster spectroscopically confirmed by the
KMOS VVVX-GalCen spectroscopic survey: a system of 14 members at
$\langle z \rangle =0.089\pm0.001$ with $\sigma_v=560^{+110}_{-120}$~km~s$^{-1}$ and
an indicative dynamical mass $M_{200}=1.9^{+0.7}_{-1.1}\times10^{14}\,M_\odot$, located at
$(l,b)=(274.05\degr,+2.85\degr)$. While the present data firmly establish the cluster core, its full spatial extent and richness remain to be determined.

The detection illustrates the complementarity of the two datasets. VVV/VVVX photometry provided 
a clear overdensity, while KMOS spectroscopy transformed the projected overdensity into a kinematically confirmed galaxy cluster. The red-clump-based extinction map described in Sect.~\ref{sec:data} shows pronounced spatial variations on scales of several arcminutes, as expected at low Galactic latitudes. Within the displayed radius of $r\sim0.25^\circ$, $A_{K_s}$ ranges from approximately $0.05$ to $0.45$~mag. At the position of the BCG, $A_{K_s}\approx0.13$~mag, coinciding with a relatively low-extinction region. While the RS candidates are
distributed throughout the field, including regions of higher local extinction, the spectroscopically confirmed members are tightly clustered around the BCG, coinciding with this local minimum in $A_{K_s}$.

The cluster lies in the direction of the Vela supercluster
\citep[VSCL;][]{kraan2017}, within the longitude range
($\ell\sim272\degr$--$278\degr$) where its two principal wall-like
structures intersect at low Galactic latitude \citep{Rajohnson2024}.
The nearer and more prominent wall lies at
$cz\sim18\,500$--$19\,000$~km~s$^{-1}$, while the second lies at
$cz\sim22\,000$~km~s$^{-1}$ \citep{kraan2017, Rajohnson2024}.
At $\langle cz \rangle = 26\,660$~km~s$^{-1}$, the new cluster
lies $\sim8\,000$~km~s$^{-1}$ beyond the first wall,
but only $\sim4\,700$~km~s$^{-1}$ (i.e.\ $\sim80$~Mpc) beyond the second. 
Therefore, we find it is a background system, rather than a member of the known VSCL structures, although
its relation to a possible continuation of the second wall (or to a more
distant structure along this sightline) remains open.

This result demonstrates the capability of the KMOS VVVX-GalCen Survey to
provide dynamical characterisations of galaxy systems behind the Galactic
plane. Forthcoming works in this series will report a second confirmed cluster
in the bulge region, examine the large-scale environment of both systems, and
characterise the properties of the \zoa\ galaxy population as a whole.

\begin{acknowledgements}
We would like to thank the Referee for constructive comments and suggestions that improved the manuscript.
This study was partially supported by Consejo Nacional de Investigaciones
Cient\'ificas y T\'ecnicas (CONICET) and Secretar\'ia de Ciencia y T\'ecnica de
la Universidad Nacional de C\'ordoba (SeCyT).
We gratefully acknowledge the use of data from the ESO Public Survey programme
IDs 179.B-2002 and 198.B-2004 obtained with the VISTA telescope. 
VVV and VVVX data are published in the ESO Science Archive under DOIs \url{https://doi.eso.org/10.18727/archive/67} and
\url{https://doi.eso.org/10.18727/archive/68}. Based on observations taken at ESO Paranal Observatory, for the KMOS VVVX-GalCen Spectroscopic Public Survey, ESO Programme ID: 116.29EF. The data are published in the ESO Science Archive Facility and are identified by the DOI: \url{https://doi.eso.org/10.18727/archive/38}.
\end{acknowledgements}

\bibliographystyle{aa}
\bibliography{ref}

\onecolumn
\begin{center}
\textbf{Authors and affiliations}
\end{center}

\noindent
E.~O.~Schmidt\textsuperscript{1,2,$\star$},
C.~Valotto\textsuperscript{1,2},
C.~Villalon\textsuperscript{1,2},
L.~D.~Baravalle\textsuperscript{1,2},
J.~L.~Nilo-Castell\'on\textsuperscript{3},
M.~V.~Alonso\textsuperscript{1,2},
M.~Soto\textsuperscript{4},
M.~G\'omez\textsuperscript{5},
D.~Minniti\textsuperscript{5,6},
W.~Weidmann\textsuperscript{2},
D.~Mast\textsuperscript{2},
C.~Obasi\textsuperscript{5},
D.~Alvarez Rocha\textsuperscript{3},
S.~Federle\textsuperscript{5},
R.~K.~Saito\textsuperscript{7},
V.~D.~Ivanov\textsuperscript{8},
M.~G.~Navarro\textsuperscript{9},
Z.~Guo\textsuperscript{10,11},
A.~Pichel\textsuperscript{12},
I.~Petralia\textsuperscript{5},
V.~Fermiano\textsuperscript{10,11},
A.~Luna\textsuperscript{13},
M.~Lad\textsuperscript{5},
J.~Osses\textsuperscript{10,11,14},
R.~Pignata\textsuperscript{2},
V.~Valdivia Rojas\textsuperscript{3},
P.~Marchant Cortes\textsuperscript{3},
E. G. Rivera\textsuperscript{3},
G.~Ferrero\textsuperscript{15,16},
N.~Masetti\textsuperscript{17,5},
J.~Borissova\textsuperscript{10},
J.~G.~ Fern\'andez-Trincado\textsuperscript{18},
C.~C\'aceres\textsuperscript{5},
F.~Nogueras-Lara\textsuperscript{19},
D.~Galdeano\textsuperscript{20},
F.~Fortuni\textsuperscript{9}.

\bigskip
\footnotesize
\noindent
\textsuperscript{1} Instituto de Astronom\'{i}a Te\'orica y Experimental (IATE-CONICET), Laprida 854, X5000BGR, C\'ordoba, Argentina\\
\textsuperscript{2} Observatorio Astron\'omico de C\'ordoba, Universidad Nacional de C\'ordoba, Laprida 854, X5000BGR, C\'ordoba, Argentina\\
\textsuperscript{3} Departamento Astronom\'ia, Universidad de La Serena, Av.\ Ra\'ul Bitran 1305, La Serena, Regi\'on de Coquimbo, Chile\\
\textsuperscript{4} Instituto de Astronom\'a y Ciencia Planetaria, Universidad de Atacama, Avenida Copayapu 485, Copiap\'o, Chile\\
\textsuperscript{5} Instituto de Astrof{\'i}sica, Departamento de F{\'i}sica y Astronom{\'i}a, Facultad de Ciencias Exactas, Universidad Andres Bello, Av. Fern{\'a}ndez Concha 700, Santiago, Chile\\
\textsuperscript{6} Vatican Observatory, V00120 Vatican City State, Italy\\
\textsuperscript{7} Departamento de F\'isica, Universidade Federal de Santa Catarina, Trindade 88040-900, Florian\'opolis-SC, Brazil\\
\textsuperscript{8} European Southern Observatory, Karl-Schwarzschild-Stra\ss{}e 2, D-85748 Garching bei M\"unchen, Germany\\
\textsuperscript{9} Istituto Nazionale di Astrofisica - Osservatorio Astronomico di Roma (INAF-OAR), Via Frascati 33, 00078, Monte Porzio Catone (RM), Italy\\
\textsuperscript{10} Instituto de F\'isica y Astronom\'ia, Universidad de Valpara\'iso, Av. Gran Breta\~na 1111, Playa Ancha, Casilla 5030, Valpara{\'i}so, Chile\\
\textsuperscript{11} Chinese Academy of Sciences South America Center for Astronomy (CASSACA), National Astronomical Observatories, CAS, Beijing 100101, China\\
\textsuperscript{12} Instituto de Astronom\'ia y F\'isica del Espacio, CONICET-UBA, Av. Int. Guiraldes 2620, C1428BNB CABA, Argentina\\
\textsuperscript{13} European Southern Observatory, Alonso de Córdova 3107, Vitacura, Santiago, Chile\\
\textsuperscript{14} Purple Mountain Observatory, Chinese Academy of Sciences, 10 Yuanhua Road, Nanjing 210023, China\\
\textsuperscript{15} Facultad de Ciencias Astron\'omicas y Geof\'isicas, Universidad Nacional de La Plata, La Plata, Paseo del Bosque S/N, B1900FWA, Argentina\\
\textsuperscript{16} Instituto de Astrof\'isica de La Plata (CONICET-UNLP), La Plata, Paseo del Bosque S/N, B1900FWA, Argentina\\
\textsuperscript{17} Istituto Nazionale di Astrofisica, Osservatorio di Astrofisica e Scienza dello Spazio di Bologna, Via Gobetti 101, 40129, Bologna, Italy\\
\textsuperscript{18} Centro de investigaci\'on en Astronom\'ia, Facultad de Ingenier\'ia, Ciencia y Tecnolog\'ia, Universidad Bernardo O'Higgins, Av. Viel 1497, Santiago, 8370993, Chile\\
\textsuperscript{19} Instituto de Astrof\'{i}sica de Andaluc\'{i}a, CSIC, Glorieta de la Astronom\'ia s/n, 18008 Granada, Spain\\
\textsuperscript{20} Gabinete de Astronom\'ia Extragal\'actica, Departamento de Geof\'isica, Astronom\'ia y F\'isica, CONICET, Facultad de Ciencias Exactas, F\'isicas y Naturales, Universidad Nacional de San Juan, Av. Ignacio de la Roza 590 (O), J5402DCS, Rivadavia, San Juan, Argentina\\

\normalsize
\blfootnote{$\star$ Corresponding author: E. O. Schmidt, \texttt{eduardo.schmidt@unc.edu.ar}}

\begin{appendix}

\section{Photometric and spectroscopic properties of the galaxies}
\label{sec:appendix}

\begin{table*}[h!]
\caption{\label{tab:cat}Photometric and spectroscopic properties of the 19 galaxy candidates identified in the photometric overdensity.}
\centering
\renewcommand{\arraystretch}{0.92}
\setlength{\tabcolsep}{5pt}
\footnotesize
\begin{tabular}{lcccccc}
\hline\hline
ID & RA & Dec & $K_{\rm s}$ & $(J-K_{\rm s})$ & $cz$ & Class \\
 & (J2000) & (J2000) & (mag) & (mag) & (km~s$^{-1}$) & \\
\hline
VVVX-J094111.90$-$490108.3 & 09:41:11.90 & $-$49:01:08.3 & 15.90 & 1.85 & \textbf{$14114 \pm 37$}  & nM  \\ 
VVVX-J094127.11$-$490146.3 & 09:41:27.11 & $-$49:01:46.3 & 15.60 & 0.94 & \textbf{$25381 \pm 37$}   & M   \\ 
VVVX-J094130.57$-$490352.1 & 09:41:30.57 & $-$49:03:52.1 & 14.55 & 1.03 & \textbf{$27464 \pm 56$}   & M   \\
VVVX-J094132.71$-$490442.1 & 09:41:32.71 & $-$49:04:42.1 & 15.15 & 1.05 & \textbf{$26791 \pm 80$}   & M   \\ 
VVVX-J094134.84$-$490349.5 & 09:41:34.84 & $-$49:03:49.5 & 15.20 & 0.92 & \textbf{$25817 \pm 34$}   & M   \\ 
VVVX-J094137.62$-$490302.5 & 09:41:37.62 & $-$49:03:02.5 & 13.19 & 1.12 & \textbf{$26665 \pm 53$}   & M$^{a}$\\ 
VVVX-J094137.79$-$490231.1 & 09:41:37.79 & $-$49:02:31.1 & 13.67 & 1.17 & \textbf{$26574 \pm 64$}   & M   \\
VVVX-J094138.01$-$490253.1 & 09:41:38.01 & $-$49:02:53.1 & 14.50 & 1.01 & \textbf{$26450 \pm 60$}   & M   \\ 
VVVX-J094141.50$-$490049.7 & 09:41:41.50 & $-$49:00:49.7 & 14.03 & 1.10 & --              & n/a \\  
VVVX-J094143.31$-$490033.7 & 09:41:43.31 & $-$49:00:33.7 & 14.17 & 1.21 & \textbf{$26727 \pm 56$}   & M   \\
VVVX-J094145.38$-$490441.7 & 09:41:45.38 & $-$49:04:41.7 & 15.53 & 1.59 & \textbf{$6236 \pm 16$}    & nM   \\  
VVVX-J094146.12$-$490431.2 & 09:41:46.12 & $-$49:04:31.2 & 15.64 & 1.06 & \textbf{$27473 \pm 22$}   & M   \\ 
VVVX-J094146.26$-$490851.0 & 09:41:46.26 & $-$49:08:51.0 & 14.74 & 1.03 & \textbf{$26827 \pm 40$}   & M   \\ 
VVVX-J094146.74$-$490952.1 & 09:41:46.74 & $-$49:09:52.1 & 13.55 & 0.99 & --              & n/a \\ 
VVVX-J094147.75$-$490652.1 & 09:41:47.75 & $-$49:06:52.1 & 14.69 & 1.05 & \textbf{$25595 \pm 20$}   & M   \\
VVVX-J094148.32$-$490930.2 & 09:41:48.32 & $-$49:09:30.2 & 14.67 & 1.00 & --              & n/a \\
VVVX-J094152.13$-$490659.7 & 09:41:52.13 & $-$49:06:59.7 & 13.28 & 1.08 & \textbf{$26660 \pm 50$}   & M   \\ 
VVVX-J094154.66$-$490848.1 & 09:41:54.66 & $-$49:08:48.1 & 13.63 & 1.02 & \textbf{$26076 \pm 41$}  & M   \\ 
VVVX-J094155.65$-$490723.4 & 09:41:55.65 & $-$49:07:23.4 & 14.69 & 1.04 & \textbf{$27122 \pm 42$}  & M   \\
\hline
\end{tabular}
\tablefoot{
The columns list the VVVX galaxy identifier, equatorial coordinates (J2000), total $K_{\rm s}$ magnitude, $(J-K_{\rm s})$ colour, barycentric radial velocity $cz$ and its uncertainty, and the adopted membership classification: M denotes spectroscopically confirmed cluster members, including the BCG (M$^{a}$). The non-members are denoted by nM, and n/a denotes candidates that were not observed.
}
\end{table*}

\end{appendix}

\end{document}